# Probing interfacial water via color-center-enabled spin magnetometry

Kang Xu[1], Kapila Elkaduwe[2,3], Rohma Khan[1,2], Sang-Jun Lee[4], Dennis Nordlund[4], Gustavo E. López[2,5], Abraham Wolcott[1,6], Daniela Pagliero[1], Nicolas Giovambattista[2,3], and Carlos A. Meriles[1,2,†]

**Abstract:** Understanding the behavior of confined water at liquid–solid interfaces is central to numerous physical, chemical, and biological processes, yet remains experimentally challenging. Here, we utilize shallow nitrogen-vacancy (NV) centers in diamond to investigate the nanoscale dynamics of interfacial water confined between the diamond surface and an overlying fluorinated oil droplet. Using NV-based nuclear magnetic resonance protocols selectively sensitive to $^1$H and $^{19}$F, we independently track water and oil near the interface under ambient conditions. Comparing opposite sides of a doubly-implanted diamond membrane — one exposed to oil, the other not — we uncover a slow, multi-day process in which the interfacial water layer is gradually depleted. This desorption appears to be driven by sustained interactions with the fluorinated oil and is supported by molecular dynamics simulations and surface-sensitive X-ray spectroscopies. Our findings provide molecular-level insight into long-timescale hydration dynamics and underscore the power of NV-NMR for probing liquid–solid heterointerfaces with chemical specificity.

**Keywords**: Liquid-solid interfaces, confined water, nitrogen-vacancy centers, nanoscale sensing.

## 1. Introduction

Interfacial water plays a crucial role in electrochemistry [1], catalysis [2], and surface science [3], influencing applications from energy conversion and oil recovery to environmental remediation and pharmaceutical formulations. At the nanoscale, interfacial water dynamics are dictated by a complex interplay of hydrophobic and hydrophilic interactions [4-6], charge accumulation [7,8], and electric double-layer formation [9,10]. Understanding these molecular-level interactions is essential yet difficult. A variety of experimental techniques have been employed to study the structure and dynamics of interfacial water, including X-ray-based methods, vibrational spectroscopies, and conventional nuclear magnetic resonance (NMR) [11-16]. X-ray scattering and absorption techniques have provided valuable insights into interfacial water density and hydrogen-bonding structures, but they typically require high-intensity radiation sources and only offer ensemble-averaged information across multiple molecular layers [11,12]. Vibrational spectroscopies, such as infrared [13], sum-frequency generation (SFG) [14], and Raman spectroscopy [15], have been widely applied to characterize interfacial O–H bonding structures. However, their sensitivity to vibrational modes often introduces ambiguities in spectral interpretation, particularly under applied electric potentials. In the same vein, conventional liquid-state NMR can probe molecular dynamics, but its application to interfacial water is severely limited by the need for bulk-phase signal averaging and insufficient spatial resolution [16].

Nitrogen-vacancy-based nuclear magnetic resonance (NV-NMR) offers a powerful, label-free approach to probe interfacial water dynamics [17-19]. Unlike other optical methods, NV-NMR detects local nuclear spin interactions with nanometer-scale resolution, enabling the quantification of diffusion and adsorption processes under ambient conditions. This technique extends the capabilities of interfacial studies by providing molecular-level insights into hydration and transport phenomena without the need for surface-specific selection rules or large-scale instrumentation such as a synchrotron facility.

Here, we utilize NV-NMR to investigate the behavior of water confined at the interface between a chemically oxidized diamond surface and perfluoropolyether (PFPE) oil. By leveraging the distinct Larmor frequencies of protons and fluorine nuclei, we selectively probe the adsorbed water layer and the PFPE oil layer via dynamical-decoupling-based protocols. Specifically, time correlation measurements allow us to study the diffusivity of adsorbed $H_2O$ molecules, much reduced at the diamond surface compared to that expected for bulk water. Comparison with a reference oil-free diamond surface over a two-week time window shows a slow but sustained impact of PFPE on the $H_2O$ dynamics, gradually reducing the water self-diffusivity. Moreover, our findings suggest this process is accompanied by water desorption, a notion we rationalize with the help of molecular dynamics simulations as well as X-ray absorption and photo-electron spectroscopies.

---

[1]Department of Physics, CUNY–The City College of New York, New York, NY 10031, USA. [2]CUNY–Graduate Center, New York, NY 10016, USA. [3]Department of Physics, CUNY–Brooklyn College of the City University of New York, Brooklyn, NY, 11210, USA. [4]Stanford Synchrotron Radiation Lightsource, SLAC National Accelerator Laboratory, Menlo Park, CA 94025, USA. [5]Department of Chemistry, CUNY–Lehman College, Bronx, NY 10468, USA. [6]Department of Chemistry, San José State University, San José, CA 95192, USA. †E-mail: cmeriles@ccny.cuny.edu

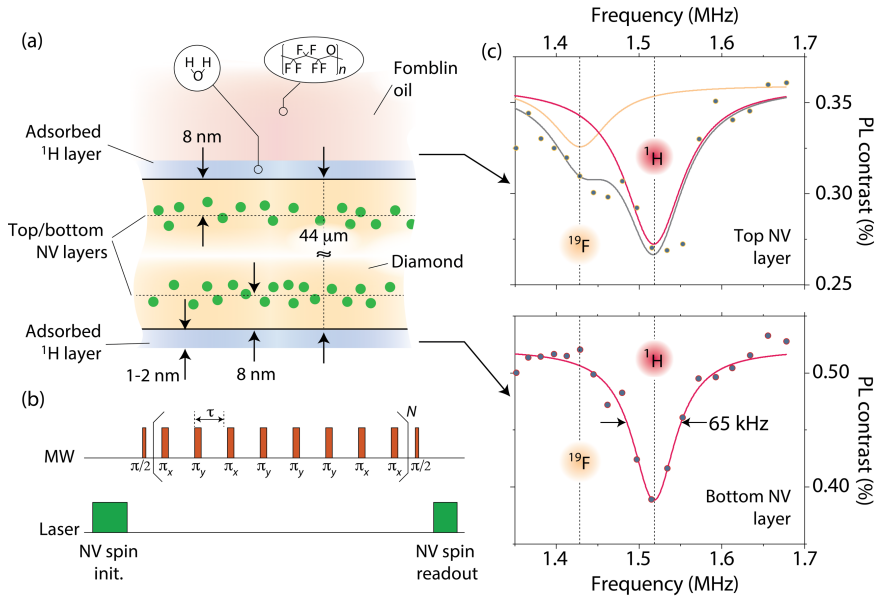

**Figure 1 | NMR spectroscopy of interfacial molecules.** (a) We use shallow NV centers created on both sides of a 44-μm-thick diamond membrane to probe nuclear spins proximal to the crystal surface. With the help of a perfluorinated oil, we spectroscopically separate molecules in the fluid from adsorbed water molecules. (b) To detect nuclear spins, we implement NV dynamical decoupling protocols of varying inter-pulse separation. (c) NV-detected NMR spectroscopy as seen on the top and bottom diamond surfaces; solid traces represent Gaussian fits and $N = 8$. We observe the $^{19}$F spin resonance exclusively on the side of the membrane exposed to oil. The magnetic field is $B = 35.7$ mT. Init: Initialization. MW: Microwave. PL: Photoluminescence.

## 2. Results

In our experiments, we utilize a 44-μm-thick diamond membrane engineered via ion implantation and annealing to host NV centers approximately 8 nm beneath the surface on both sides (Fig. 1a). The μm-size of the membrane ensures both NV sets fall within the working distance of the microscope objective, a convenient feature allowing us to compare the response of one NV set against the other (see below). A thin layer of adsorbed water, typically a few nanometers thick, spontaneously forms on both surfaces within minutes upon exposure to ambient humidity[20-23]. In our case, this adsorption process is likely initiated by the acid cleaning stage, leveraged here to induce partial surface oxygenation and hence facilitate the formation of negatively charged NVs (see Experimental Section).

To detect nanoscale NMR signals from surface nuclei, we first employ a dynamical decoupling (DD) scheme in the form of an XY8-$N$ pulse train resonant with one of the NV ground state spin transitions[24,25]. This protocol comprises a series of microwave (MW) pulses separated by a time interval τ (Fig. 1b), which collectively renders the NV sensitive to external AC magnetic fields within a narrow frequency window centered at $1/2\tau$. As the inter-pulse spacing varies, a characteristic NMR signal emerges when the filter frequency matches the nuclear spin Larmor precession frequency under the applied static magnetic field.

In our measurements, we deposit a small amount of PFPE oil (FOMBLIN Y LVAC Grade 25/6) on only one of the membrane surfaces, leaving the other in pristine form ("top" and "bottom" sides of the membrane, respectively). PFPE oil is a liquid lubricant comprising polymeric chains of perfluoro-alkyl groups connected by ether linkages. It contains no $^1$H but provides an abundance of $^{19}$F nuclei. Because the gyromagnetic ratios of $^1$H (42.576 MHz·T$^{-1}$ in water) and $^{19}$F (40.078 MHz·T$^{-1}$) are different, their resonance signals can be separately resolved. Figure 1c presents the NV-NMR spectra obtained immediately after casting a PFPE drop on the top diamond surface. We find that NVs both in the oil-coated and oil-free surfaces exhibit comparable photoluminescence (PL) dips at the $^1$H Larmor frequency. On the other hand, only the oil-coated side exhibits a $^{19}$F resonance dip, demonstrating the ability of our system to selectively detect water and PFPE molecules. Interestingly, the $^1$H NMR signal remains the dominant feature in the recorded spectrum, suggesting that the thickness of the adsorbed water layer — presumably displacing the PFPE molecules away from the surface — is substantial; taking the different densities for H and F nuclei into account, we estimate not less than 1 – 2 nm, although an accurate determination of the thickness is unwarranted.

The ~100 MHz spectral separation between $^1$H and $^{19}$F resonances at 35.7 mT allows us to probe the dynamics of water and PFPE diffusion selectively. To this end, we implement a spin correlation protocol, i.e., a sequence designed to produce a signal proportional to the product of the nuclear-spin-induced phases picked up by the NV during two sensing windows separated by a variable time interval[26-28] (Fig. 2a). In the present case, each sensing window comprises an XY8-4 sequence, providing a filter bandwidth of approximately 1/16 of the target nuclear spin Larmor frequency. Crucially, the NV response resulting from a correlation protocol vanishes gradually as molecules leaving the NV sensing volume are replaced by others statistically polarized; the consequence is that the decay rate extracted from the signal envelope serves as a measure of molecular self-diffusion selective to the liquid under inspection[18,27,29].

Figure 2b shows the results for PFPE, where we observe a damped sinusoidal centered at the ¹⁹F Larmor frequency and featuring a long coherence time $T_2^{(F)}$, exceeding 60 μs. The decay approximately follows an exponential function, though a long-lived tail seems to persist, possibly arising from adsorbed molecules at the interface undergoing rotational motion without translational diffusion[27], and/or from modifications in the shape of the auto-correlation function due to the restricted diffusion geometry created by the diamond surface[29]. The FFT spectrum of the ¹⁹F signal is centered at 1.429 MHz with a linewidth of approximately 15 kHz, that is likely limited by the 10 kHz frequency resolution limit stemming from our 100 μs measurement time.

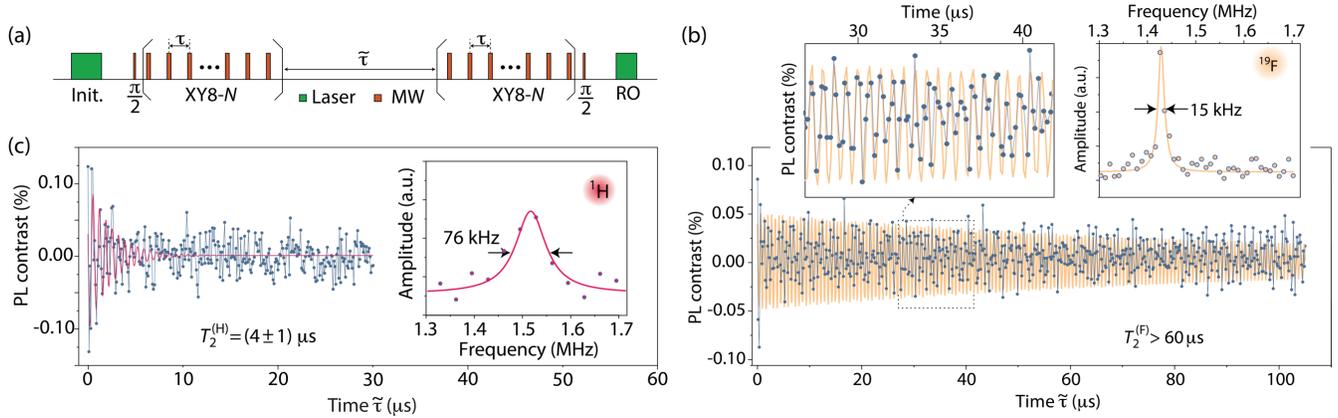

**Figure 2 | ¹H and ¹⁹F NMR correlation spectroscopy.** (a) Experimental protocol. We correlate the phase picked up by the NVs before and after an evolution interval of variable duration $\tilde{\tau}$. (b) (Main) Spin correlation signal as a function of $\tilde{\tau}$. With the interpulse spacing $\tau$ set to half the ¹⁹F spin Larmor period, we make detection selectively sensitive to ¹⁹F spins. The solid yellow trace is a fit to an exponentially damped sinusoid at the ¹⁹F Larmor frequency. (Upper left insert) Zoomed view of the ¹⁹F correlation signal throughout the time interval within the dashed box in the main figure; we measure a decoherence time $T_2^{(F)}$ exceeding 60 μs. (Upper right insert) Fourier transform of the signal in in the main plot showing a peak at the expected ¹⁹F Larmor frequency (1.429 MHz at the applied magnetic field). (c) (Main) Same as above but upon setting $\tau$ to half the ¹H Larmor period. We find a much shorter-lived signal, decaying on a characteristic time $T_2^{(H)}$ of order 4 μs. (Insert) Fourier transform of the signal in the main plot yielding a resonance peak at the expected ¹H frequency. In (b) and (c), we set $N = 8$. RO: Readout.

Since the sensing radius $d$ scales with the NV depth[30], we can leverage the above measurements to estimate the PFPE diffusion constant from the approximate formula $d_{PFPE}^2 \sim 2\delta T_2^{(F)} D_{PFPE}$, where $\delta = 2 - 3$ is a unitless parameter representing the effective dimensionality of the molecular motion; in the limit of free three-dimensional diffusion and assuming the adsorbed water layer is ~2 nm thick, we set $\delta = 3$ and $d_{PFPE} \approx 10$ nm to obtain $D_{PFPE} \sim 0.3 \times 10^{-12}$ m² s⁻¹, comparable to the bulk value ($0.5 \times 10^{-12}$ m² s⁻¹, see Experimental Section). The discrepancy could stem from a slightly reduced PFPE mobility near the interface with the adsorbed water layer, but could also be a consequence of the geometric restriction imposed by the surface (we obtain nearly the bulk diffusion constant for $\delta = 2$).

Figure 2c shows an extension of the above measurements, where we configure the spin correlation protocol to selectively probe protons. Here, we find a much shorter correlation time $T_2^{(H)} \approx 4$ μs, consistent with the 76 kHz linewidth of the ¹H Larmor resonance observable at 1.518 MHz after Fourier transform. From the equivalent relation, $d_{H2O}^2 \sim 2\delta T_2^{(H)} D_{H2O}$, we derive $D_{H2O} \approx 4 \times 10^{-12}$ m² s⁻¹ for the water self-diffusion coefficient, where we used $d_{H2O} \approx 8$ nm and set $\delta = 2$ to incorporate the near-two-dimensional nature of the adsorbed water motion. The value range we derive for $D_{H2O}$ is peculiar: On the one hand, it is orders of magnitude smaller than that corresponding to bulk water[31] ($2.3 \times 10^{-9}$ m² s⁻¹), implying the motion of interfacial H₂O molecules is strongly hindered. The reduction we observe, however, is not as large as reported recently[17], where ¹H spin correlation measurements yielded signals lasting for several tens of microseconds (corresponding to diffusion coefficients comparable to those seen here for PFPE). This variability could be a combination of surface termination and/or charge accumulation, invoked to explain the long coherence lifetime of H₂O protons confined to a nanochannel structure[18]. Interestingly, we observe a nearly identical response in the oil-free side of the membrane (not shown for brevity) suggesting the effect is largely due to interactions with the diamond surface.

Unexpectedly, the presence of PFPE has a slow but sustained effect on the water layer. This is shown in Fig. 3a, where we display the NV-NMR spectrum for the PFPE-coated after 10 days of contact. The dynamic decoupling spectrum from the top diamond surface reveals a pronounced decrease in the ¹H resonance signal — both in amplitude and spectral width — compared to that measured after just one day, thus hinting at reduced water diffusivity, possibly due to increased confinement (e.g., through the formation of patches, see below) and/or reduced thickness. At the same time, the $^{19}$F resonance grows stronger and sharper, eventually dominating the spectrum, suggesting that PFPE molecules progressively develop a stronger and more localized coupling with near-surface NV centers likely due to increased proximity. In stark contrast, Fig. 3b shows that the NV-NMR spectrum from the oil-free, bottom side of the diamond membrane remains virtually unchanged over the same 10-day period.

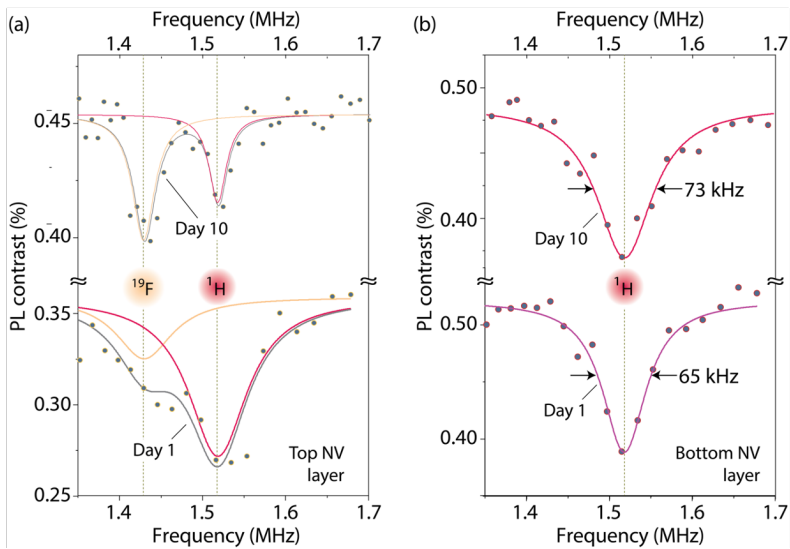

**Figure 3 | Time evolution of the adsorbed layer.** (a) NMR spectroscopy of the top surface using the protocol in Fig. 1b. Comparison of the spectra after a 10-day exposure to PFPE oil (upper data set) shows a much-depleted $^1$H resonance relative to that obtained at earlier times (lower data set, here reproduced for clarity from Fig. 1c); we also observe the growth and narrowing of the $^{19}$F resonance. Combined, these observations suggest oil-induced erosion of the adsorbed layer and altered mobility of near-surface oil molecules. (b) Same as above but for protons adsorbed on the lower diamond surface. We find virtually no change over the same 10-day time interval.

To gain insight into the molecular mechanisms underlying the experimental findings in this work, we also perform molecular dynamics (MD) simulations aimed at visualizing how oil molecules interact with water at the diamond surface over time. Since the force fields for hydrocarbons are more extensively validated and reliable than those for fluorinated species, we use dodecane as a representative oil phase. Our goal is to provide a qualitative — rather than quantitative — understanding of the behavior of water at the interface with diamond in the presence of large hydrophobic molecules.

The upper-left panel of Fig. 4 shows the initial configuration of our model system. At $t = 0$, the OH-terminated diamond substrate is fully covered by a ~1 nm-thick water film containing 5372 water molecules. Above it, we add a thick, 3000-molecule dodecane ($C_{12}H_{26}$) film (see Supplementary Material (SM), Section I.1 for MD simulations details). Snapshots taken at $t = 2, 4,$ and 8 ns illustrate the progressive interfacial water/dodecane rearrangements at $T = 300$ K. During the first 1-2 ns, the water film remains adsorbed on the diamond surface, preventing dodecane from reaching the solid. However, at about $t = 2$ ns, the water film breaks down; as time proceeds, the water film evolves into a water droplet that remains in contact with the diamond surface and surrounded by dodecane (see snapshots for $t = 4, 8$ ns). To explore the influence of surface chemistry, we also perform MD simulations using diamond surfaces with O- and H-terminations. In all cases, we find that the water film evolves into a water droplet surrounded by dodecane but in contact with the diamond surface (SM, Section I.2), which points to a process largely independent of the diamond termination. We also investigated the impact of temperature on our MD simulations: As shown in the SM, Section I.5, increasing the temperature does not affect our conclusions but it speeds up the transformation of the water film into a droplet.

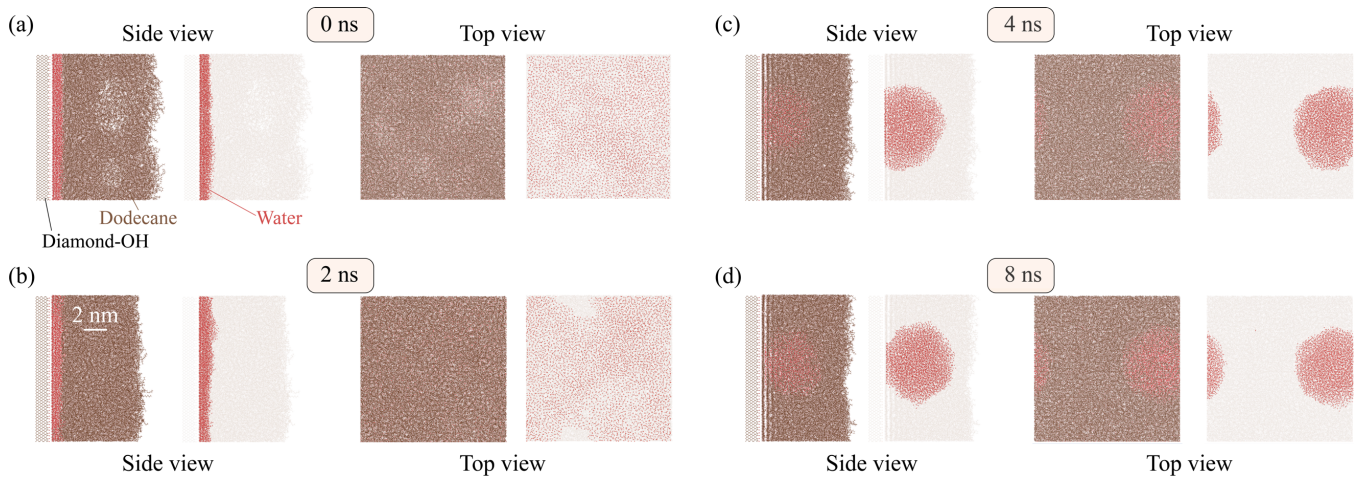

**Figure 4 | Dodecane-driven displacement of interfacial water as seen by MD simulations.** The modeled system contains 5372 water molecules (red) and 3000 dodecane molecules (brown) placed on an OH-terminated diamond surface at $T = 300$ K. Configurations are taken at four different times during the MD simulations, $t = 0, 2, 4,$ and 8 ns (a through d, respectively). At each time, the snapshots provide side and top views; we also include in each case water-selective views for clarity. At $t = 0$, the system is prepared so water forms a uniform ~1 nm-thick film covering the diamond interface, with dodecane laying on top of the water film. The water film and dodecane remain visible for at least 2 ns. As the simulation proceeds, the dodecane molecules gradually migrate toward the diamond surface, while the water film evolves into a droplet that remains in contact with the hydroxylated diamond surface (the water droplet shown at $t = 8$ ns is split due to periodic boundary conditions).

The results in Figure 4 indicate that large oil molecules, such as dodecane, can displace thin water films (~1 nm-thick) from the diamond surface. However, the time scales for water removal in the MD simulations are orders of magnitude shorter than in the experiment. Thicker water films may bring the simulation time scales closer to the experimental times (SM, Section I.4). Yet, other phenomena may also be at play. Recent experimental studies have shown that shallow NV centers (and other surface defects) under laser excitation can inject carriers into water layers formed on the diamond surface[32-34]. To emulate this effect, we add 1352 excess anions (fluoride) to the water film shown in Figure 4. To maintain charge neutrality in the system, we add a compensating positive charge distributed uniformly over the diamond surface (see SM, Section I.1, and Ref. [18] for details). While dodecane still migrates toward the diamond surface, a stable water monolayer persists while the remaining displaced water molecules reorganize into bubble-like domains (SM, Section I.3). These results suggest that interfacial charges help stabilize the water layer and alter the restructuring dynamics. Lastly, we also note that dodecane is a highly nonpolar hydrocarbon, whereas PFPE is chemically more complex, featuring fluorine atoms with significantly larger negative partial charges. These differences may give rise to interfacial behaviors that differ somewhat from those observed with dodecane. Nonetheless, the important and non-trivial point from our MD simulation is that large hydrophobic molecules, such as dodecane, can indeed displace water molecules from the diamond surface with different surface terminations.

To further probe the chemical nature of the residual species, we performed synchrotron-based X-ray absorption spectroscopy (XAS) on PFPE-coated CVD diamond samples (see Experimental Section). In the C1$s$ edge (Fig. 5a), PFPE exposure introduces strong σ*(C–F) transitions within the range 291–300 eV, while attenuating the native diamond features, consistent with coverage of the oxidized diamond surface by fluorinated oil. F1$s$ XAS reveals two characteristic σ*(C–F) peaks at ~688.6 and ~691.5 eV, assigned to secondary and primary alkyl-fluorides, respectively (Fig. 5b). O1$s$ spectra show a similar trend: The native π*(C=O) signal is reduced, while the σ*(C–O) region intensifies due to PFPE's ether functionalities (Fig. 5c).

To investigate the chemical stability of fluorinated species on the diamond surface, we performed X-ray photoelectron spectroscopy (XPS) on the same samples. Even after repeated solvent rinsing and piranha cleaning, the F1$s$ peak remained clearly detectable (SM, Section II), indicating the robust retention — or possible chemical bonding — of PFPE-derived species at the interfaces[35]. Together, these data indicate that PFPE forms a stable, chemically distinct interfacial layer on the diamond, consistent with suppressed water accessibility and the long-lived $^{19}$F signal detected by NV-NMR.

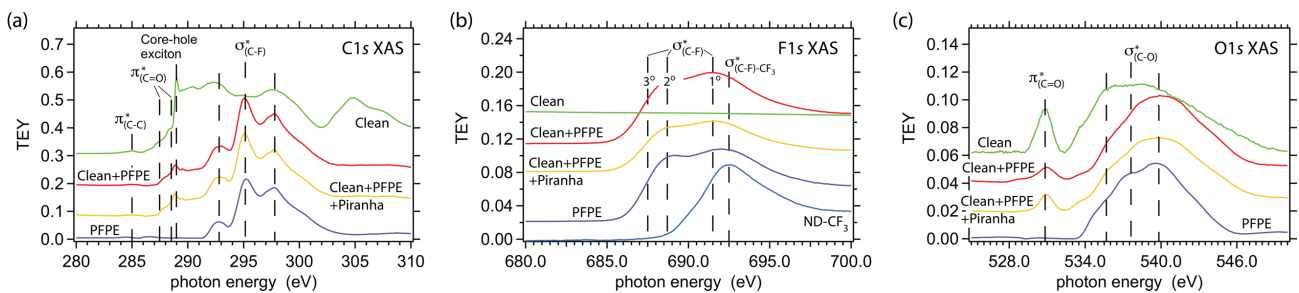

**Figure S5 | XAS analysis of diamond surfaces under different treatments.** (a) C1$s$ XAS spectra showing "clean" diamond and after exposure to PFPE leading to attenuation of the diamond electronic structure including the core-hole exciton at 289.0 eV and the 2$^{nd}$ absolute bandgap. Diamond and PFPE have almost a 4-eV differential in absorption onset with PFPE absorption from 292.5-297.6 eV for the σ*(C-O) and σ*(C-F) resonances. (b) F1$s$ XAS spectra displaying persistent σ*(C-F) features at ~689 eV and ~692 eV, even after piranha cleaning, indicating robust retention of fluorinated species after typical oxidation treatments. (c) O1$s$ XAS spectra revealing reduction in oxygen functionalities due to diamond, with reduced π*(C=O) at 530.8 eV and enhanced σ*(C-O) signals after PFPE exposure. In (a) through (c), traces have been displaced vertically for clarity; see Experimental Section for additional details. TEY: Total electron yield.

## 3. Conclusion

In this study, we employed NV-center-based quantum sensing to investigate the nanoscale dynamics of interfacial water confined at a diamond-oil interface. By selectively probing $^{1}$H and $^{19}$F nuclei using correlation spectroscopy, we identified distinct diffusion behaviors for water and PFPE molecules, and uncovered a slow, multi-day process in which oil progressively displaces the interfacial water layer. Molecular dynamics simulations revealed that this rearrangement is largely independent of the diamond surface termination; further, our simulations show that the time scale of the process — typically in the ns range — can be altered in the presence of injected charges and/or thicker adsorbed water layers. Complementary XAS and XPS measurements confirm the persistent presence of fluorinated species at the diamond surface after PFPE exposure, supporting the interpretation of the NV-NMR observations.

Our results establish NV-NMR as a powerful technique for tracking time-dependent interfacial processes with molecular specificity and nanometer-scale resolution. The ability to observe molecular exchange at buried solid-liquid interfaces during biologically and technologically relevant timescales opens new possibilities for studying dynamic systems in electrochemistry, lubrication, and membrane science. More broadly, this approach provides a versatile platform for exploring heterogeneous fluid interfaces and for informing the design of interface-sensitive materials and devices.

## 4. Experimental Section

*Sample preparation and characteristics*: We employed a customized [100] electronic-grade diamond membrane supplied by Qnami AG, with a thickness of 44 µm. A 6 µm deep trench was fabricated on one surface via reactive ion etching (RIE), serving as a microfluidic channel for liquid confinement. To create nitrogen-vacancy (NV) centers, we performed $^{14}$N ion implantation at 2.5 keV with a dose of 1×10$^{13}$ ions/cm², resulting in a nitrogen-rich layer approximately 8 nm deep, with a ~2 nm spread. NV center formation was induced through a standard high-temperature annealing protocol, with an end temperature of 1000 °C. The sample underwent cleaning and oxygen termination using a tri-acid mixture (HNO$_3$: H$_2$SO$_4$: HCl) followed by Piranha solution (H$_2$SO$_4$: H$_2$O$_2$) to remove organic contaminants and ensure a well-defined surface chemistry.

*Experimental setup*: We made use of a custom-built confocal fluorescence microscope integrated with a Tabor Proteus arbitrary waveform generator (AWG) to establish a quantum sensing platform based on diamond color centers. The confocal microscopy setup uses a 532 nm laser (approximately 300 µW) to excite NV-center fluorescence through a high numerical aperture objective, forming a diffraction-limited, ~0.4-µm-radius excitation spot. The NV fluorescence is collected through a dichroic mirror and filtered by a 635 nm long-pass optical filter before being fiber-coupled into an avalanche photodiode (APD) for sensitive photon counting. The focusing depth of the objective can be precisely adjusted using a piezo-controlled stage, enabling independent addressing of NV center ensembles on either surface of the diamond membrane ensuring spatial selectivity and accurate control during measurements. Spin manipulation and quantum sensing experiments are implemented using the Tabor Proteus AWG, which features a 9 GHz sampling rate. Custom-programmed sequences allow us to perform diverse quantum sensing protocols with sub-nanosecond timing resolution, ensuring high-fidelity control of NV spin states.

*Correlation spectroscopy*: The NV correlation protocol consists of two multi-pulse XY8-*N* sequences, separated by a variable interval $\tilde{\tau}$, allowing for free nuclear spin evolution. The two DD segments at the beginning and end of the sequence, with a fixed pulse interval $\tau$ matching half the nuclear Larmor period, act as band-pass filters centered at either the $^{1}$H or $^{19}$F resonance, depending on the target nucleus. To probe the long-lasting nuclear spin coherence, the phases of the last $\pi/2$ pulses in the two segments are shifted by ±90° relative to the first $\pi/2$ pulse. This adjustment stores the accumulated phase into the NV spin magnetization, effectively extending the measurable signal lifetime up to the NV $T_1$ relaxation limit, thereby enabling nuclear spin evolution measurements in the millisecond regime. By varying $\tilde{\tau}$, the coherent phase evolution of the nuclear-spin-induced magnetic field can be mapped, allowing the Larmor precession to be observed as a free induction decay (FID) signal. Given the long nuclear spin coherence times typical in liquids (where internuclear dipolar couplings average out due to fast molecular motion), the correlated coherence buildup signal primarily reflects molecules that remain within the detection volume during the measurement. Consequently, the decay of the signal as a function of $\tilde{\tau}$ inherently captures the molecular spatial diffusion time.

*Estimation of self-diffusion coefficient*: In our experiment, the PFPE oil is FOMBLIN Y LVAC Grade 25/6 purchased from SPI Supplies. The bulk self-diffusion coefficient, $D_{\text{PFPE}}$, was estimated via the Stokes-Einstein equation:

$$D_{\text{PFPE}} = \frac{k_B \cdot T}{6\pi \cdot \eta \cdot r},$$

where $k_B$ denotes the Boltzmann constant, $T$ is the temperature, $\eta$ is the dynamic viscosity, and $r$ is the effective hydrodynamic radius of the diffusing molecule. From the data sheet, the kinematic viscosity ($\nu$) of the liquid was given as 276 cSt (276 × 10$^{-6}$ m²/s) and the density ($\rho$) as 1.90 g/cm³ (1900 kg/m³). The dynamic viscosity ($\eta$) was calculated as $\eta = \nu \cdot \rho = 0.5244$ Pa·s. Its average molecular weight is 3300 amu, the molecular radius ($r$) was estimated from the molecular weight (M) and density ($\rho$), which is about 8.83 × 10$^{-10}$ m. Finally, the self-diffusion coefficient ($D_{\text{PFPE}}$) of the liquid at 20 °C was estimated to be 4.63 × 10$^{-13}$ m² s$^{-1}$.

*Analysis of XAS measurements*: To investigate the diamond-PFPE interface we performed synchrotron-based X-ray absorption spectroscopy (XAS). After chemical oxidation, 100-terminated CVD diamond contains ketone, ether and *sp*2-like (Pandey-chain) groups[36]. At the C1*s* edge, "clean" diamond shows 2 unique diamond features, the core-hole exciton and 2nd band gap at 289.0 eV and 302 eV, respectively, with additional pre-edge features from 285-288 eV[37] (Fig. 5a). In contrast, C1*s* absorption in PFPE begins at 291 eV, 6 eV above the diamond absorption; the spectra also reveal a series of σ*(C–F) resonances from ~291-298 eV. In the absence of prior data, we assign these latter peaks to the tertiary, secondary, and primary alkyl-fluorides. Post PFPE deposition, the C1$s$ diamond signal is suppressed by ~90%, the core-hole exciton is barely observable, and the spectrum is overtaken by the PFPE overlayer. Suppression of diamond signals by an overlayer is similar to silica coated nanodiamonds probed by C1$s$ and is consistent with the inelastic mean free path of secondary Auger electrons[38].

F1$s$ spectra of PFPE-diamond and 1 control illustrate the blue shift in σ*(C-F) as the local bonding environment transitions from tertiary (3°), to secondary (2°), to primary (1°) alklyl fluorides (Fig. 5b). PFPE oil and PFPE-diamond is dominated by the σ*(C-F) peaks at 688.6 eV and 691.5 eV due to the secondary and primary bonding environments, respectively. As a control, we are also including the ND-CF$_3$ spectrum from a 3,3,3-trifluoro-propyl-trimethyoxysilane decorated silica nanodiamond sample, which reflects the primary alkyl fluorides at 692.5 eV[38]. We also measured a sample of fluorinated high-pressure high-temperature nanodiamond (ND-F, not shown for brevity) that contains tertiary fluorides with the lowest σ*(C-F) at 687.5 eV, and is ~1.5 eV lower in energy than fluorine-terminated [100] diamond[39]. Lastly, O1$s$ spectra show a convolution of π*(C=O) signal from ketones at ~530.8 eV which is then overtaken by the σ*(C-O) ethers linkages of PFPE oil above 534 eV (Fig 5c).

## Acknowledgments

We thank Alexander Pines for early discussions and help procuring the diamond membranes. The authors also thank Artur Lozovoi for valuable assistance with the experimental setup. K.X., K.E., N.G., and C.A.M acknowledge support from the National Science Foundation through NSF-2223461; D.P. acknowledges funding from the National Science Foundation via grant NSF-2203904. A.W. acknowledges funding from the National Science Foundation via grant NSF-2112550. All authors acknowledge access to the facilities and research infrastructure of the NSF CREST IDEALS, grant number NSF-2112550. This work used computational resources at San Diego Supercomputer Center (SDSC) through allocation grant CHE240016 from the Advanced Cyberinfrastructure Coordination Ecosystem: Services & Support (ACCESS) program, which is supported by NSF Grants 2138259, 2138286, 2138307, 2137603, and 2138296.

## Conflict of interest

The authors declare that they have no conflict of interest.

## Data availability

The data that support the findings of this study are available from the corresponding author upon reasonable request.

## Code availability

All source codes for data analysis and numerical modeling used in this study are available from the corresponding author upon reasonable request.

## Author contributions

K.X., and C.A.M. conceived the experiments. K.X., R.K., A.W., and D.P. conducted experiments, whereas K.E., G.L., and N.G. carried out the numerical modeling of interfacial water dynamics. A.W., S.J.L., and D.N. carried out the XAS and XPS measurements. All authors analyzed the data; K.X. and C.A.M. wrote the manuscript with input from all authors.

## Competing interests

The authors declare no competing interests.

## Correspondence

Correspondence and requests for materials should be addressed to C.A.M.

# Supplementary Material for

# Probing interfacial water via color-center-enabled spin magnetometry

Kang Xu[1], Kapila Elkaduwe[2,3], Rohma Khan[1,2], Sang-Jun Lee[4], Dennis Nordlund[4], Gustavo E. López[2,5], Abraham Wolcott[1,6], Daniela Pagliero[1], Nicolas Giovambattista[2,3], and Carlos A. Meriles[1,2,†]

[1]*Department of Physics, CUNY–The City College of New York, New York, NY 10031, USA.*
[2]*CUNY-Graduate Center, New York, NY 10016, USA.*
[3]*Department of Physics, CUNY–Brooklyn College of the City University of New York, Brooklyn, NY, 11210, USA.*
[4]*Stanford Synchrotron Radiation Lightsource, SLAC National Accelerator Laboratory, Menlo Park, CA 94025, USA.*
[5]*Department of Chemistry, CUNY–Lehman College, Bronx, NY 10468, USA.*
[6]*Department of Chemistry, San José State University, San José, CA 95192, USA.*

[†]Corresponding author: cmeriles@ccny.cuny.edu.

**I. Molecular dynamics simulations**

To further understand the interfacial dynamics underlying the experimentally observed erosion of the water film, we conducted molecular dynamics (MD) simulations of systems containing ~1 nm-, ~2 nm-, and ~3 nm-thick water films sandwiched between a diamond surface and a ~7 nm-thick dodecane film. The dodecane film is composed of 3000 dodecane molecules, and the water films contain $N=5372$, $10467$ and $16135$ molecules. Three different diamond surfaces are considered with the following surface terminations: hydrogen (H), oxygen (O), and hydroxyl groups (OH). We use similar geometries as in our previous work [1]. Briefly, the diamond surfaces have an area $A = \Delta y \cdot \Delta x$ with $\Delta x, \Delta y = 13.137 - 13.197$ nm depending on the surface considered. The diamond crystal thickness is $\Delta z = 1.151 - 1.226$ nm (depending on the diamond surface termination). Increasing the value of $\Delta z$ does not affect our results because the additional C atoms would be located at least 1.15 nm apart from any water molecule in the film and the corresponding C-water interactions would be null (the Lennard-Jones interactions between the water molecules and the diamond inner C atoms are short-ranged, and are evaluated within a cutoff distance $r = 1.10$ nm) [1]. Water molecules are represented using the TIP4P/2005 water model [2] while the dodecane molecules are modelled using the charmm36 force field [3]. The diamond model surfaces are provided in Ref. [1].

All MD simulations are performed at constant temperature and volume. Periodic boundary conditions apply to all three directions with $L_x=\Delta x$, $L_y=\Delta y$, and $L_z=30.00$ nm (note that $L_z$ is much larger than the diamond thickness $\Delta z$ and the water/dodecane film thickness; this extra empty space is included in order to minimize any potential effect that may be due to the periodic boundary conditions along the z-direction). All computer simulations are performed using the GROMACS software package (2022.6-spack) [4]. The system temperature is maintained at a constant value using a Nose-Hoover thermostat with a time constant of 0.5 ps [5,6]. Coulombic interactions are evaluated using the particle mesh Ewald technique [7] with a Fourier spacing of 0.12 nm. For both the Lennard-Jones (LJ) and the real space Coulomb interactions, a cutoff $r = 1.10$ nm is used. To control the water OH covalent bonds lengths, we use the LINCS (Linear Constraint Solver) algorithm. MD simulations are performed for 8 ns which is long compared with the relaxation time of bulk TIP4P/2005 water ($\tau$~50–100 ps) at $T = 300$ K and $P = 0.1$ MPa [8] (under these conditions, it takes <100 ps for the mean-square displacement of the water molecules to reach 1 nm$^2$). The simulation time step is $dt = 1$ fs. All figures of MD simulations are prepared with VESTA 3 software package [9].

**I.1 System preparation**

The upper-left panel of Figure 4 in the main text shows an example of the starting configurations used in our MD simulations. The initial water film (of thicknesses ~1, ~2, and ~3 nm) is extracted from an independent MD simulation where a ~6.5 nm-thick water film ($N = 33628$) is placed on a hydroxylated diamond surface. The ~6.5 nm-thick water film on diamond is simulated for 4 ns, and the last configuration is used to extract three water films of reduced thicknesses ~1, ~2, and ~3 nm containing, respectively, $N=5372$, $10467$ and $16135$ water molecules. The three water films are then treated as follows.

A) Each water film is placed on a diamond surface terminated with either H, O, or OH groups. The water film-diamond surface systems are then equilibrated for 4 ns at 300 K. In all cases, we confirm that the water film is stable on the corresponding diamond surface.

B) For each of the equilibrated systems obtained in (A), we add a ~7 nm-thick dodecane film, in turn, obtained from an independent MD simulation of dodecane (3000 molecules) covering a hydroxylated diamond surface at $T=300$ K. The dodecane film is placed at approximately 1-1.3 nm away from the water film surface. The diamond-water-dodecane system evolves for an additional 40 ps in a MD simulation where the water molecules are kept immobile. This allows the dodecane molecules to approach the water film and adjust to the water film interface (see, e.g., the upper-left snapshot of Figure 4 in the main manuscript).

C) The diamond-water-dodecane systems obtained in (C) are then used as the starting configuration of MD simulations performed for 8 ns.

The same protocol described above was used for the systems with injected charges (fluoride); see Section I.3). In these cases, however, the initial ~6 nm-thick water film contains $N=29090$ water molecules (instead of $N=33628$), and 1352 fluoride anions. In addition, to maintain the system charge-neutral, a compensating positive charge is added to the diamond surfaces (see below). During equilibration, the 1352 fluoride anions move towards the diamond surface (independent of the surface termination) and remain covered by a thick film of water. This allows us to extract water films of thicknesses ~1, ~2, and ~3 nm containing 1352 fluoride anions with identical number of water molecules ($N=5372$, $10467$ and $16135$) used in the systems that do not contain fluoride. The fluoride anions are modeled using the Madrid2019 force field [10] where each anion has an effective charge $q = -0.85$ $e$. Hence, to make the system charge-neutral, we also add a total positive charge $N_q |q|$ to the diamond crystal. This charge is distributed evenly over the C monolayer of the diamond crystal that is located closest to the water film. This C monolayer contains $N_c$ atoms ($N_c = 2 N_q = 2704$ in the case of diamond crystals passivated with H and OH groups; $N_c = 2738$ for the diamond terminated with O groups). The specific distribution of the positive charge within the diamond crystal (e.g., all the positive charge added to one C monolayer, or distributed over all the C atoms in the diamond crystal) does not affect our results.

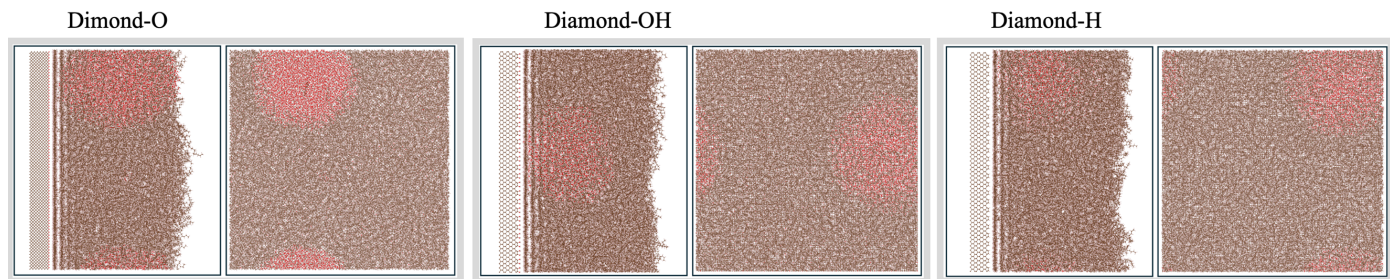

**Figure S1 | Snapshots from MD simulations of water, dodecane, and diamond with different surface terminations.** All systems contain 5372 water molecules and 3000 dodecane molecules. Panels (left-to-right) are obtained at $T$=300 K and $t$=8 ns, with diamond surfaces passivated with oxygen (O), hydroxyl (OH), or hydrogen (H) terminations, respectively. In each case, we include the side and top view of the system. Water molecules are shown in red, and dodecane molecules in brown. In all three cases, a water droplet is observed in contact with the diamond surface, and embedded within the dodecane film, indicating that water is gradually displaced from the diamond interface over time.

### I.2 Impact of surface termination

Figure S1 shows snapshots from our MD simulations of the diamond-water-dodecane systems for the cases where the diamond surfaces with different surface termination; in each panel, we include the side and top views of the system (left- and right-hand sides, respectively). In all cases, the water film which initially covers the whole diamond surface is gradually displaced by dodecane. This results in the formation of a water droplet embedded within the dodecane volume but attached to the diamond surface (as observed in Figure 4 of the main text). The consistent behavior observed across the H-, O-, and OH-terminated diamond surfaces indicates that the water displacement in the presence of dodecane does not depend on the surface chemistry. However, our results show a slight variation in the water contact angle with the diamond termination considered.

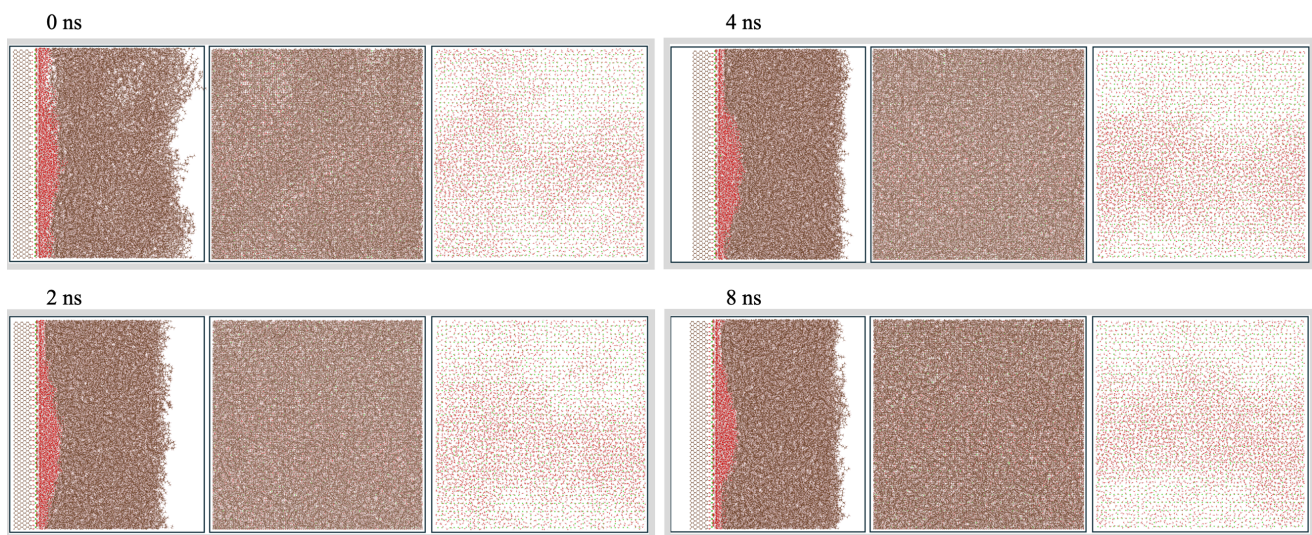

**Figure S2 | Snapshots from MD simulations of water, dodecane, and injected charges (fluoride) on diamond.** The system consists of 5372 water molecules (red), 3000 dodecane molecules (brown), and 1352 fluoride anions (green) in contact with an OH-terminated diamond crystal. Snapshots are shown at $t$=0, 2, 4, and 8 ns. For each case, we show the side and top view of the system (left and middle panels), and the top view of the water subsystem. The compensating (positive) charge added to the diamond crystal stabilizes a water monolayer diamond surface, which remains throughout the MD simulation. Meanwhile, the remaining water molecules reorganize forming bubble-like domains, under the dodecane film.

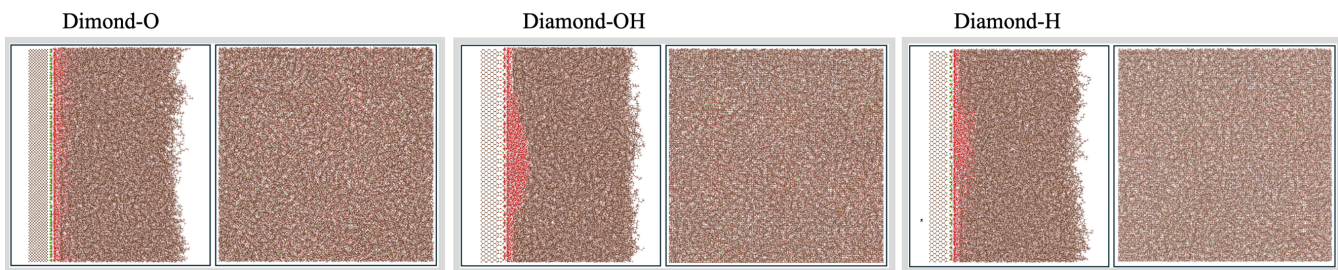

**Figure S3 | Snapshots from MD simulations of water, dodecane, and injected charges (fluoride) on diamond with different surface terminations.** All systems contain 5372 water molecules (red), 3000 dodecane molecules (brown), and 1352 fluoride anions (green). Snapshots (left-to-right) are obtained at $T$=300 K and $t$=8 ns, with diamond surfaces passivated with oxygen (O), hydroxyl (OH), or hydrogen (H) terminations, respectively. Each panel includes the side and top views of the system. In all three cases, a water monolayer remains adsorbed on the diamond surface; a water-droplet-like volume forms under the dodecane film.

### I.3. Impact of charge injection

To model charge injection in the water-dodecane-diamond systems studied, we also perform MD simulations with additional 1352 fluoride anions. The negative charge added to the system due to the inclusion of fluoride is compensated by adding the corresponding positive charge to the outermost C layer of the diamond surface (see Sec. I.1 and Ref. [1]). This model system is meant to provide insights into the effects of charge injection in the water films due to the laser excitations of the NV centers. As shown in Figure S2, the presence of injected charges (fluoride) leads to the stabilization of a water monolayer adsorbed at the diamond surface that cannot be removed by the dodecane molecules. Remarkably, the remaining water molecules still form localized bubble-like domains under the dodecane film. These results hold for all three surface terminations of diamond (Figure S3).

### I.4. Impact of water film thickness

The thickness of the water film considered plays a relevant role. To show this, we perform MD simulations on water and dodecane in contact with diamond ($T$=300 K). Our MD simulations show that water films that are approximately >2nm-thick remain on the diamond surface and are not removed by the surrounding dodecane molecules (within the 8-ns simulation window), regardless of the surface termination (Figure S4, middle and right panels). We further test the water film stability by performing additional MD simulations of the ~2 nm-thick water film at higher temperatures, $T$=325, 350 K. In all cases, we find that the water film remains stable during the 8ns-long MD simulation.

### I.5. Impact of temperature

To investigate the impact of temperature on our results, additional MD simulations were conducted at $T$=325, 350 K with the ~1 nm-thick water film in the presence of dodecane, and in contact with the OH-passivated diamond surface. MD simulations are performed both with and without injected charges (fluoride). In all cases, increasing the temperature from

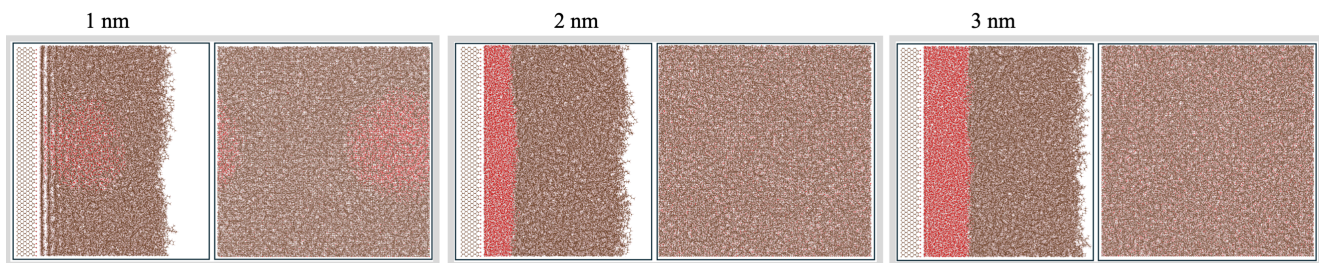

**Figure S4 | Snapshots from MD simulations of water, dodecane, and diamond with water films of different thickness.** Snapshots (side and top views) are obtained at $T$=300 K and $t$=8 ns. The water films are (left-to-right) ~1 nm-, ~2 nm-, and ~3 nm-thick and contain $N$=5372, 10467, and 16135 molecules, respectively. In all cases, the dodecane film contains 3000 molecules and the diamond surface is terminated with hydroxyl groups (OH). Water and dodecane molecules are shown in red and brown, respectively. Only for the ~1 nm-thick water film, the water molecules reorganize and form a droplet embedded within the dodecane film, indicating that water is gradually displaced from the diamond interface over time (the ~2 nm- and ~3 nm-thick water films remain stable within the simulated time (8 ns)).

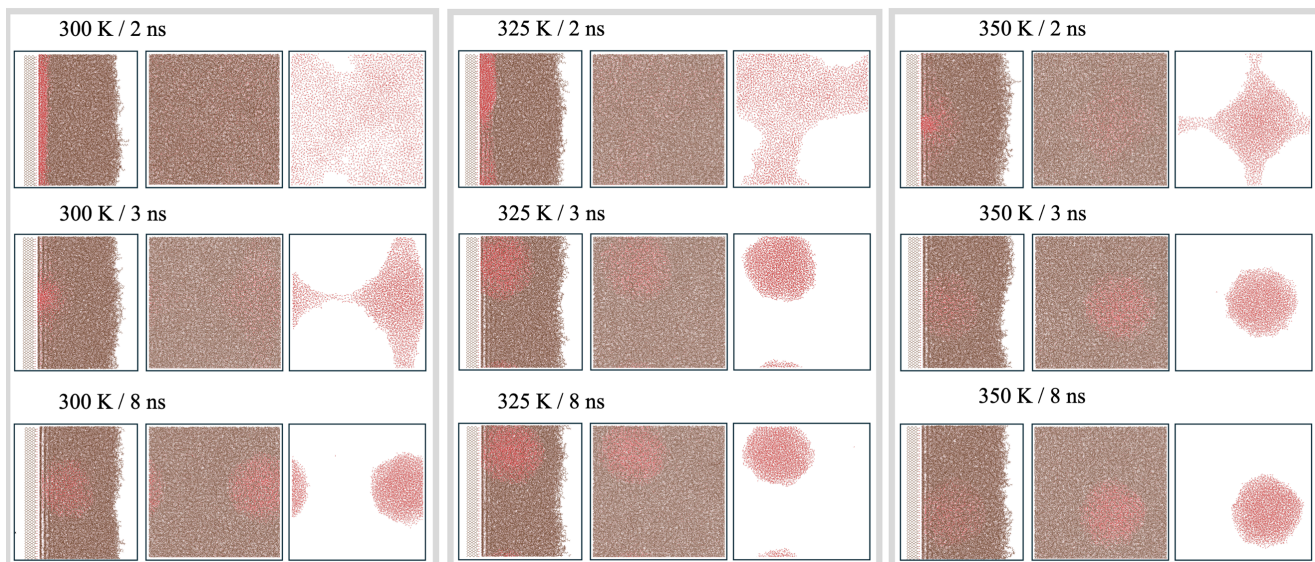

**Figure S5 | Snapshots from MD simulations of water, dodecane, and diamond at different temperatures.** The system is composed of a ~1 nm-thick water film ($N$=5372, red) sandwiched by an OH-terminated diamond surface and a thick dodecane film (3000 molecules, brown). Left, middle, and right column correspond to MD simulations performed at $T$=300, 325, and 350 K (the system at $T$=300 K is the same system included in Figure 4 in the main text). For each temperature and time, we include the side and top views of the system, and the top view of the water subsystem. In all cases, the water film is displaced by dodecane, and a water droplet forms on the diamond surface. The effect of increasing the temperature is to speed up the water droplet formation.

$T$=300 K to $T$=325, 350 K does not affect our conclusions. Specifically, in the absence of injected charges (fluoride), the water film evolves into a droplet at all the temperatures studied (Figure S5). Increasing the temperature speeds up the water droplet formation. Similarly, in the presence of fluoride (injected charges), a water monolayer remains adsorbed at the diamond surface at all temperatures, while a droplet-like water volume forms under the dodecane film (Figure S6).

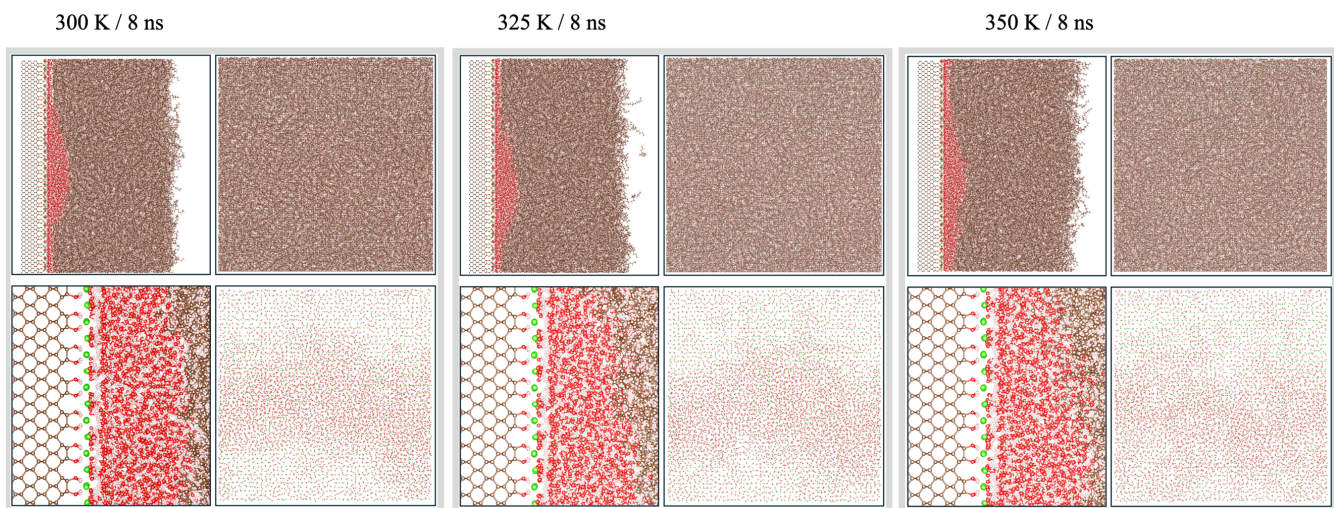

**Figure S6 | Snapshots from MD simulations of water, dodecane, and injected charges (fluoride) on diamond at different temperatures.** The system simulated is the same shown in Figure S2. Left, middle, and right columns are snapshots from MD simulations performed at $T$=300, 325, and 350 K respectively ($t$=8 ns). Each column shows the side (upper-left) and top views of the system (upper-right); the top view of the water subsystem is included in the bottom-right panel. The bottom-left panel is a zoomed out view of the diamond-water interface. The increase in temperature does not impose significant changes to the system within the 8 ns simulation time window. A water monolayer remains adsorbed on the diamond surface and a droplet-like water volume develops under the dodecane film.

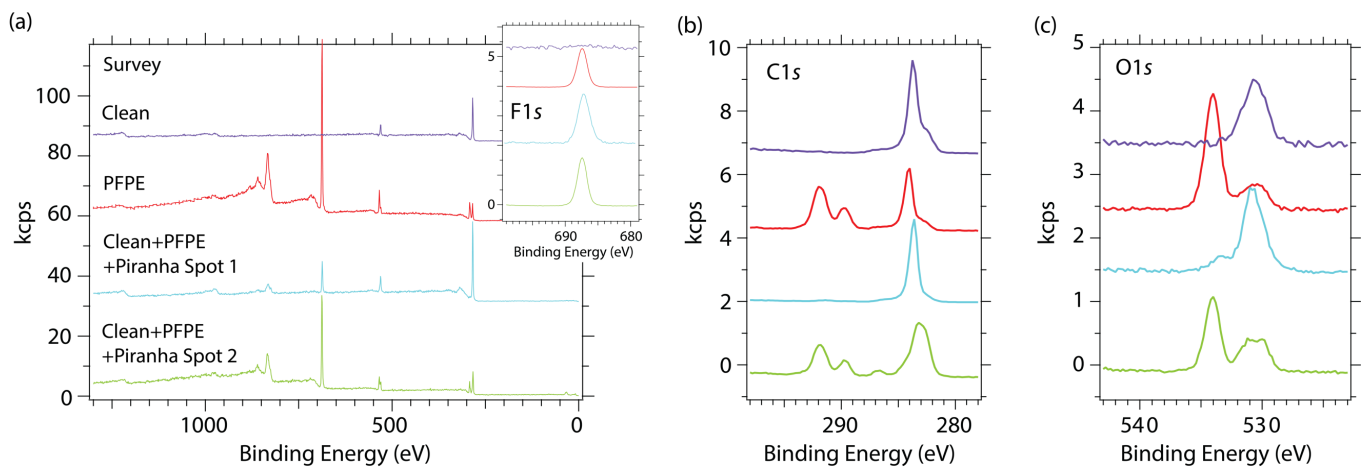

**Figure S7 | XPS characterization of diamond surfaces before and after PFPE oil exposure and piranha cleaning.** (a) Survey spectra for the diamond upon four different treatments, showing global changes in elemental composition. Insert: high-resolution F1$s$ spectra. From top to bottom: no detectable F for the pristine sample; strong F signal after PFPE exposure; residual F1$s$ signal remains detectable after piranha cleaning at two different locations, indicating robust retention of fluorinated species. (b, c) Same as in (a) but for the C1$s$ and O1$s$ energies, respectively. The color coding in the plots follows that in (a). kcps: kilo counts per second.

**II. XPS Analysis of PFPE residue on diamond surfaces**

To assess the robustness and potential chemical bonding of fluorinated species from PFPE oil at the diamond surface, we performed X-ray photoelectron spectroscopy (XPS) measurements on a CVD diamond sample (BD27) after various treatments. BD27 is a CVD-grown single-crystal diamond substrate from Element Six, implanted with nitrogen under similar energy and depth conditions as those used in our NV-containing membranes. The surface treatments included: (1) piranha-cleaned pristine diamond, (2) diamond coated with PFPE oil, and (3) PFPE-coated diamond after a second round of piranha cleaning.

XPS measurements were carried out with a monochromatic Al K$\alpha$ (1486.6 eV) source under ultrahigh vacuum (~$10^{-9}$ Torr). Survey scans were followed by high-resolution scans over selected regions (C1$s$, O1$s$, and F1$s$). As shown in Figure S7, the pristine diamond surface exhibits no detectable fluorine signal, whereas a pronounced F1$s$ peak emerges following PFPE oil exposure. Notably, this fluorine signal remains clearly detectable even after an extensive cleaning protocol consisting of three sequential rinses with acetone, IPA, and DI water, followed by ~30 minutes of piranha cleaning. At both examined positions (Spot 1 and Spot 2), the F1$s$ peak intensity decreases slightly but remains substantial, indicating robust surface retention of fluorinated species. Additional high-resolution C1$s$ and O1$s$ spectra reveal bonding changes and oxygen content variations associated with PFPE adsorption and oxidative processing.